\documentclass[preprint,showpacs,preprintnumbers,amsmath,amssymb]{revtex4-1}
\usepackage{graphicx}
\usepackage{verbatim}
\usepackage{epstopdf}
\usepackage[usenames]{color}
\usepackage{graphics}  
\usepackage{dcolumn}  
\usepackage{bm}          
\begin{document}
\input{epsf}

\preprint{APS/12three-QED}

\title{Few-body semiclassical approach to nucleon transfer and emission reactions}

\author{Renat A. Sultanov\footnote{rasultanov@stcloudstate.edu;\ \ r.sultanov2@yahoo.com}
and D. Guster\footnote{dcguster@stcloudstate.edu} }

\affiliation{Department of Information Systems \& BCRL at St. Cloud State University, 
Integrated Science and Engineering Laboratory Facility (ISELF), St. Cloud, MN 56301-4498, USA}

\date{\today}

\begin{abstract}
A three-body semiclassical model is proposed to describe the nucleon transfer and
emission reactions in a heavy-ion collision. In this model the two heavy particles, i.e. nuclear cores
A$_1(Z_{A_1}, M_{A_1})$ and A$_2(Z_{A_2}, M_{A_2})$,
move along classical trajectories $\vec R_1(t)$ and $\vec R_2(t)$ respectively,
while the dynamics of the lighter neutron, n, is considered from
a quantum mechanical point of view. Here, $M_i$ are the nucleon masses and $Z_i$ are the Coulomb charges of
the heavy nuclei ($i=1,2$). A Faddeev-type semiclassical formulation using realistic paired
nuclear-nuclear potentials is applied so that all three channels (elastic, rearrangement and break-up)
are described in an unified manner. In order to solve these time-dependent equations the Faddeev
components of the total three-body wave-function are expanded 
in terms of the input and output channel target eigenfunctions.
In the special case when the nuclear cores are identical (A$_1 \equiv$ A$_2$) and the
two-level approximation in the expansion over target functions
the time-dependent semiclassical Faddeev equations are resolved in an explicit way.
To determine the realistic $\vec R_1(t)$ and $\vec R_2(t)$ trajectories of the nuclear cores
a self-consistent approach based on the Feynman path integral theory is applied.
\end{abstract}
\maketitle

\section{Introduction}
When trying to describe nuclear collisions, compound and halo nuclei, or, for instance,
complex nuclear fusion reactions, few-body models are extremely useful and can play a very important 
role in the field \cite{khalili2002,orland2013}. For example,
in works \cite{hiyama2013,hiyama2012,deltuva2010} the authors widely use various
few-body models of complex nuclei
for numerical computation of different systems and nuclear reactions.
In an older paper \cite{fedorov96} a detailed few-body approach has
been developed for calculation of an important problem in nuclear astrophysics, namely the
first two 0$^+$ levels in the nucleus $^{12}$C which was considered
as a model three $\alpha$-particle system \cite{thompson,ishikawa,fedorov2001}.
Specifically, well known three-body
Faddeev equations \cite{faddeev60} were used in \cite{fedorov96,fedorov2001}. Further, in the case
of heavy-ion collisions a three-body semiclassical model has been introduced in \cite{revai85}. 
Once again a Faddeev-type formulation was utilized featuring single-term separable (non-local) potentials
between particles. For solution of the few-body equations a simplified semiclassical model was applied,
where heavier nuclear cores of the system followed along straight-line classical trajectories. 
Therefore, the resulting model featuring the semiclassical Faddeev equations become
a set of coupled time-dependent integral equations.
More generally, in the case of the heavy ion collisions \cite{li2008}
various semiclassical models have been
formulated and applied, see for example \cite{malfliet74,fujita76,phillips77}. However, these approaches
mainly used simple straight-line model trajectories \cite{brink72}. Also, there are other
interesting and important problems in the field of heavy-ion collisions such as neutron and a charge
transfer and emission reactions \cite{deep1980}. In the framework of the Faddeev approach these
channels can be treated in a unified manner.

Nontheless,  in \cite{fujita76} an interesting attempt has been made to expand this process.
In this work the author tried to
apply a semiclassical Pechukas formalism \cite{pechukas69} to obtain more realistic classical
trajectories of heavy nuclear particles. Pechukas's method was originally developed for atomic,
molecular collisions and chemical reactions. This theory expands on Feynman's interpretation of
quantum-mechanics, which is based on path integrals \cite{feynman65}.
Usually, semiclassical methods and models, such as \cite{pechukas69}, 
allow us to gain even deeper insight into different few-body or
many-body physical systems. 
They also enable us to introduce even more realistic
classical trajectories of heavier particles in models, i.e. to take into account
quantum-mechanical corrections in a self-consistent manner \cite{pechukas69}. 

Generally speaking, such a
combination of few-body models and methods together with semiclassical models, where
the dynamics of heavier particles can be separated from the dynamics of lighter particles, seems to be quite
useful. The same approach has already been developed and widely used in some problems of chemical physics for
molecular dynamics \cite{billing87} and even for the description of many body systems, see for example \cite{kapral1999}.

In the current work we develop a semiclassical model for a few-body treatment of neutron transfer and emission reactions
in heavy ion collisions at different impact energies. The (A$_1,n)$ + A$_2$ system is shown in Fig. 1, where A$_1$ and
A$_2$ are the heavy nuclear cores which move along classical trajectories $\vec R_1(t)$ and $\vec R_2(t)$. The inter-distance
vector $\vec R_{12}(t)=\vec R_1(t)-\vec R_2(2)$ is also shown in Fig. 1 together with the coordinate $\vec r$ of the third particle, i.e.
neutron n, $\vec \rho$ is the impact parameter, $\vec v_1 (\vec v_1\ ^{\prime})$ and $\vec v_2 (\vec v_2\ ^{\prime})$ 
are the initial (final) velocities of the heavy particles, $O$ is the
center-of-mass of the three-body system. The semiclassical model of a time-dependent set of Faddeev equations is used.
However, in contrast to Revai's approach \cite{revai85} we formulate the model with the use of two {\it local} (realistic)
paired nuclear-nuclear potentials between the A$_1$ particle and n and between A$_2$ and n.
The heavy nuclei A$_1(Z_{A_1}, M_{A_1})$
and A$_2(Z_{A_2}, M_{A_2})$ move along realistic classical trajectories $\vec R_1(t)$
and $\vec R_2(t)$, while the motion of the relatively light neutron n $(m_n \ll M_{A_i})$ in their nuclear
fields is treated from a quantum mechanical point of view. In this model
the heavy particles can move along complex Coulomb trajectories.
This problem is particularly important for lower energy collisions and small impact parameters, i.e. at $\rho\approx 0$:
when the use of simple straight-line trajectories does not provide an appropriate approximation \cite{sult2003}.
In this work we employ a self-consistent Pechukas
method \cite{pechukas69} which provides a proper way to determine the true trajectories of
the heavy classical particles \cite{fujita76,sult2003}.
In the next section we will delineate our semiclassical formalism. The self-consistent Pechukas approach
is also explained and applied to the three-body semiclassical system as shown in Fig. 1.

\section{Semiclassical model}
In this section a few-body semiclassical model for a single neutron, n, transition
from one heavy center to another and n-emission process when the particle reaches the continuous
spectrum is presented. In order to describe these processes in a unified way the
few-body Faddeev equation approach is applied. To solve these equations a modified close coupling method is
used in this work \cite{fb_my}. This method provides a set of coupled time-dependent differential equations with 
unknown expansion coefficients.

\subsection{Time-dependent few-body Faddeev equations}
The time-dependent integral differential
Faddeev equations \cite{faddeev60} can be delineated:
\begin{equation}
\left (i \hbar \frac{\partial}{\partial t} - H_0 - V_{jk}\right)
\vert \Psi_l \rangle = V_{jk} \left ( \vert \Psi_j \rangle +
\vert \Psi_k \rangle \right),
\label{eq:lfadd}
\end{equation}
here $H_0$ is the kinetic energy operator
of the three particles:
\begin{equation}
H_0 = -\frac{\hbar^2}{2\mu_{jk}} \Delta_{\vec r_{jk}} +
\frac{\hbar^2}{2M_l} \Delta_{\vec R_{l}},
\end{equation}
$\vec r_{jk}$ and $\vec R_{l}$ are the Jacobi coordinates,
$\mu_{jk}$ and $M_l$ the corresponding reduced masses,
$V_{jk}$ the two-body potentials.
As mentioned above we consider the third particle
(neutron, electron or muon) to be the light one, i.e.:
\begin{equation}
\frac{m_n}{M_{A_1}} \ll 1, \hspace{1cm} \frac{m_n}{M_{A_2}}
\ll 1.\label{eq:condm}
\end{equation}
Then, the heavy particles 1 and 2
can be considered as moving along
classical trajectories $\vec R_1(t)$ and $\vec R_2(t)$.
For the treatment of this situation and description of the light
particle dynamics we use instead of three coupled, time-dependent
Faddeev equations just two Faddeev-like equations \cite{sult94,sult97}:
\begin{equation}
\left (i \hbar \frac{\partial}{\partial t} - \frac{p_{r}^2}
{2m_n} - V_{13}(\vec x)\right)
\Psi_1(\vec r, \vec R(t), t)  =
V_{13}(\vec x)
\Psi_2(\vec r, \vec R(t), t),
\label{eq:fadd11}
\end{equation}
\begin{equation}
\left (i \hbar \frac{\partial}{\partial t} - \frac{p_{r}^2}
{2m_n} - V_{23}(\vec y)\right)
\Psi_2(\vec r, \vec R(t), t) =
V_{23}(\vec y)
\Psi_1(\vec r, \vec R(t), t).
\label{eq:fadd}
\end{equation}
Here, $ \vec R(t) $ is the relative vector between particles
A$_1$ and A$_2$, where the time dependence is determined according to classical mechanics.
The motion of the light particle 3 (n - neutron) is treated quantum
mechanically, $\vec p_{r} = \hbar \vec \nabla_r/i$
is the momentum operator corresponding to
the relative variable $\vec r$
between third particle ${\it n}$ and the center of mass of
particles A$_1$ and A$_2$.
The relative vectors in the subsystems  (13) and (23) are denoted by
$\vec x$ and $\vec y$, respectively.
To solve these equations (\ref{eq:fadd}), we expand the wave function
components
$\Psi_k(\vec r, \vec R(t), t)$ into the solutions
$\Phi_n^{k3}(\vec r, \vec R(t), t)$ of the respective
subsystem's Schr\" odinger   
equation:
\begin{equation}
\left(i \hbar \frac{\partial}{\partial t} - \frac{p^2_r}{2 m_n}
 - V_{k3}
(\vec r - \vec R_k(t)) \right)
\Phi_n^{k3}(\vec r, \vec R_k(t), t) = 0\;.
\label{eq:fi}
\end{equation}
That is, we can write 
\begin{equation}
\Psi_k(\vec r, \vec R(t), t) = \left (\sum + \int \right )_n
C_n^{k}(\vec R(t), t)
\Phi_n^{k3}(\vec r, \vec R(t), t), \;
\label{eq:expan}
\end{equation}
the summation (integration)
runs accross the whole discrete and continuous
spectrum. For a constant velocity $\dot {\vec R}_k(t) = \vec v_k$
one finds \cite{sult94,sult97}
\begin{equation}
\Phi_n^{k3}(\vec r, \vec R(t), t) =
e^{im_n \vec v_k \vec r/\hbar - i(E_n^{k3}
+ \frac{m_n}{2} v^2_k)t/\hbar} \varphi_n^{k3}
(\vec r - \vec R_k(t))\;,
\label{eq:trans}
\end{equation}
the functions $\varphi_n^{k3}$ as being given by
\begin{equation}
\left( - \frac{\hbar^2}{2m_n} \, \Delta_{\vec x}
+ V_{k3}(\vec x)\right)
\varphi_n^{k3}(\vec x) = E_n^{k3} \varphi_n^{k3}(\vec x)\;.
\end{equation}
 
Inserting the expansion (\ref{eq:expan}) into (\ref{eq:fadd}),
we obtain for the coefficients $C_n^{k}$
a set of coupled equations:
\begin{equation}
i\hbar\frac{\partial C_n^{1}(\vec R(t), t)}{\partial t}
= \left(\sum + \int \right)_m {\cal W}_{nm}^{12}(R(t), t)
\gamma_{nm}^{12} (t) C_{m}^{2}(\vec R(t), t),
\label{eq:maineqw}
\end{equation}
\begin{equation}
i\hbar\frac{\partial C_m^{2}(\vec R(t), t)}{\partial t}
= \left(\sum + \int \right)_n {\cal W}_{mn}^{21}(R(t), t)
\gamma_{nm}^{12*} (t) C_{n}^{1}(\vec R(t), t),
\label{eq:maineq}
\end{equation}
where
\begin{equation}
\gamma_{nm}^{(jk)}(t) = e^{i(E_n^{j3} - E_m^{k3})t/\hbar}\ (j \neq k = 1, 2).
\label{eq:ingr}
\end{equation}

The matrix elements ${\cal W}^{jk}_{nm}(R(t), t)$
are obtained by integrating the potentials in Eq. (\ref{eq:fadd}) between 
the channel functions (\ref{eq:trans}),
\begin{eqnarray}
{\cal W}_{nm}^{jk}(R(t), t) = \Bigl<
e^{im_n \vec v_j \vec r / \hbar - i\frac{m_n}{2}v^2_jt/\hbar}
\varphi^{j3}_n(\vec r-\vec R_j(t)) \Big|
V_{j3}(\vec r - \vec R_j(t))
\Big| \nonumber \\
e^{im_n \vec v_k \vec r / \hbar - i\frac{m_n}{2}v^2_kt/\hbar}
\varphi^{k3}_m(\vec r - \vec R_k(t)) \Bigr>.
\label{eq:wmatel}
\end{eqnarray}
The equations (\ref{eq:fadd}) are then to be solved under the initial conditions
\begin{equation}
\Psi_1(\vec r, \vec R(t), t)
\mathop{\mbox{\large$\sim$}}\limits_{t \rightarrow - \infty}
\Phi_{1s}^{13}(\vec r, \vec R(t),t),\ 
\Psi_2(\vec r, \vec R(t), t)
\mathop{\mbox{\large$\sim$}}\limits_{t \rightarrow - \infty} 0,
\label{eq:condpsi}
\end{equation}
which implies that for the coefficients $C_n^{j}(\vec R(t), t)$:
\begin{equation}
C_n^{1}(\vec R(t), t)
\mathop{\mbox{\large$\sim$}}\limits_{t 
\rightarrow - \infty} \delta_{n1},\ C_n^{2}(\vec R(t), t)
\mathop{\mbox{\large$\sim$}}\limits_{t \rightarrow - \infty} 0.
\label{eq:cond}
\end{equation}
For reactions at low energies the relative nuclear velocities are
practically zero in the respective unities. The exponential
factor in eq. (\ref{eq:trans}), hence, can
be replaced by the unit and the matrix elements
(\ref{eq:wmatel}) which simplifies to: 
\begin{equation}
{\cal W}_{nm}^{jk} (R(t)) \enspace = \enspace \int d^3r
\varphi_n^{j3^*}(\vec r - \vec R_j(t))
\,V_{j3}(\vec r - \vec R_j(t))
\,\varphi_m^{k3}(\vec r - \vec R_k(t))\;.
\label{eq:wmatel2}
\end{equation}
In order to obtain the capture probabilities
$\vert C_n^{2}(t \sim \infty)\vert^2$
we have to solve the system of coupled ordinary differential equations
(\ref{eq:maineqw})-(\ref{eq:maineq}).
Note that its ingredients and initial conditions are described in
(\ref{eq:ingr}), (\ref{eq:cond}) and (\ref{eq:wmatel2}).
When solving the resulting coupled set of equations for the
expansion coefficients, it is observed that its solutions
$C_n^{k}(\vec R(t), t)$ tend toward an asymptotic
value $C^k_n (\rho)$  which depends, of course, on the
impact parameter $\rho$.

The elastic and transfer semiclassical cross sections of the
three-particle collisions are \cite{pechukas69}:
\begin{equation}
\left(\frac{d\sigma}{d\Omega}\right)_{el} = \left(
\frac{d\sigma}{d\Omega}\right)_{cl}
\vert C_1^{1} (\rho) - 1\vert^2\;,
\label{eq:cross1}
\end{equation}
and
\begin{equation}
\left(\frac{d\sigma}{d\Omega}\right)_{tr} =
\left( \frac{d\sigma}{d\Omega}\right)_{cl}
\vert C_n^{2} (\rho) \vert^2\; , \label{eq:cross2}
\end{equation}
respectively, where $\left(d\sigma / d\Omega\right)_{cl}$
is the cross section of the classical scattering of
A$_1$ and A$_2$ which are heavy nuclear cores.
For the break-up channel of the reaction one can delineate:
\begin{equation}
W_{b-up}(\vec k_0, \rho)  =
\Big | < \vec k_0(t)\Big |\Psi_1(\vec r, \vec R(t), t) + \Psi_2(\vec r, \vec R(t), t) \Bigr > \Big |^2_{t\rightarrow\infty},
\label{eq:break1}
\end{equation}
where $W_{b-up}$ is the neutron emission probability and
$\big | \vec k_0(t)\bigr >$ is its wave function in the continuous spectrum:
\begin{equation}
\Big | \vec k_0(t)\Bigr > = \frac{e^{i(\vec k_0\vec r-E_0t)/\hbar}}
{\sqrt{(2\pi)^3}},
\label{eq:break2}
\end{equation}
that is:
\begin{eqnarray}
W_{b-up}(\vec k_0, \rho) & = & \frac{1}{(2\pi)^3} \biggr|\int dr^3
e^{-i/\hbar(\vec k_0\vec r-E_0t)}
\left (\sum + \int \right )_n
C_n^{1}(\rho)\Phi_n^{13}(\vec r, \vec R(t), t) \nonumber \\
& + & \int dr^3e^{-i/\hbar(\vec k_0\vec r-E_0t)}\left (\sum + \int \right )_m
C_m^{2}(\rho)\Phi_m^{23}(\vec r, \vec R(t), t) \biggr|^2,
\label{eq:break3}
\end{eqnarray}
finally one can obtain the following formula for the three-body break-up, i.e. neutron emission process:
\begin{eqnarray}
W_{b-up}(\vec k_0, \rho)& = &\frac{1}{(2\pi)^3}
\biggr|\left (\sum + \int \right )_n
C_n^{1}(\rho) e^{iE_0t/\hbar - i\frac{m_n}{2} v^2_1t/\hbar}
\int dr^3 e^{-i\vec k_0 \vec r} 
e^{i/\hbar m_n \vec v_1 \vec r/\hbar}\nonumber \\
&\times& \varphi_n^{13}(\vec r - \vec R_1(t))
+\left (\sum + \int \right )_n C_n^{2}(\rho) e^{iE_0t/\hbar - i\frac{m_n}{2} v^2_2t/\hbar}\nonumber \\
&\times& 
\int dr^3 e^{-i\vec k_0 \vec r}e^{i/\hbar m_n \vec v_2 \vec r/\hbar}\varphi_n^{23} (\vec r - \vec R_2(t))\biggr|^2.
\label{eq:emission}
\end{eqnarray}
The triple-differential cross section of this process is:
\begin{equation}
\left( \frac{d^3\sigma}{k_0^2 dk_0 d^2\Omega}\right)_{b-up} = \left( \frac{d\sigma}{d\Omega}\right)_{cl}
\vert W_{b-up}(\vec k_0, \rho)\vert^2.
\label{eq:cross3}
\end{equation}

\subsection{Application of Pechukas's self-consistent approach}
To obtain the true trajectories of the heavy classical particles or nuclear cores
A$_1(Z_{A_1},M_{A_1})$ and A$_2(Z_{A_2},M_{A_2})$
one can employ the Pechukas self-consistent method \cite{pechukas69}
based on the Feynman path-integral theory \cite{feynman65}.
In accordance with the method a reduced propagator
containing exact information about the reaction $\beta \rightarrow 
\alpha$ can be written using continual integration
\begin{equation}
{\it G}_{\alpha \beta}(\vec R_2t_2\vert \vec R_1t_1) =
\int_{\vec R_1t_1}^{\vec R_2t_2} D[\vec R(t)]
e^{iS_0[\vec R(t)] / \hbar} T_{\alpha \beta}[\vec R(t)]\;,
\end{equation}
where $S_0[\vec R(t)]$ is the classical action of the heavy particle
moving along $\vec R(t)$;
$T_{\alpha \beta}[\vec R(t)]$ are the
transition amplitudes used for
finding a quantum particle at $t_2$ in the state $\vert \alpha \rangle$
if at $t_1$ it was in the state $\vert \beta \rangle$. Obviously
$T_{\alpha \beta}$ is related with the model which was discussed above and
determined from Eqs.(\ref{eq:maineqw}) - (\ref{eq:cond}).
Now one should impose a
proper limit for the scattering or transfer problem,
i. e. $t\rightarrow \infty$, $D[\vec R(t)]$ is the measure of continual integration.
The time-dependent behaviour of the amplitudes $T_{\alpha \beta}$ is
determined by the Hamiltonian $h(t)$ \cite{pechukas69}:
\begin{equation}
h(t) = -\frac{\hbar^2}{2m_n} \triangle_{\vec r} +
V_{13}(\vert \vec r - \vec R_1^t\vert) +
V_{23}(\vert \vec r - \vec R_2^t\vert) + {\cal U}(R_{12}^t).
\end{equation}
In accordance with the self-consistent method \cite{pechukas69} a basic variational principle is:
\begin{equation}
\delta (S_0[\vec R(t)] + \hbar Im\ln T_{\alpha \beta}[\vec R(t)]) = 0.
\label{eq:vari}
\end{equation}
Variation of Eq.(\ref{eq:vari}) gives the Newton equations
for the dynamics of classical particles in the effective
potential field ${\cal V}(\vec R(t))$. It takes into account an interplay between the
classical and quantum degrees of freedom in the semiclassical system, i.e.
quantum-mechanical corrections from the third particle n \cite{pechukas69}:
\begin{equation}
M \frac{d^2 \vec R(t)}{dt^2} + \vec \nabla_R {\cal V}(\vec R(t)) = 0,
\label{eq:class}
\end{equation}
where
\begin{equation}
{\cal V}(\vec R(t)) = {\it Re}
\frac{\langle \alpha(t, t'')\vert h(t)\vert
\beta (t, t') \rangle}{\langle \alpha(t, t'')\vert \beta(t, t') \rangle}.
\end{equation}
Here, $\vert \alpha(t, t'') \rangle$ and $\vert \beta(t, t') \rangle$
are two solutions of the time-dependent Schr\" odinger equation with
different boundary conditions \cite{pechukas69}:
\begin{equation}
i / \hbar \frac{\partial}{\partial t} \vert \alpha(t, t'') \rangle =
h(t) \vert \alpha(t, t'') \rangle, \ \vert \alpha(t'', t'')\rangle = \vert \alpha \rangle,
\end{equation}
and
\begin{equation}
i / \hbar \frac{\partial}{\partial t} \vert \beta(t, t') \rangle =
h(t) \vert \beta(t, t') \rangle,\ \vert \beta(t', t')
\rangle = \vert \beta \rangle.
\end{equation}
Therefore
\begin{equation}
T_{\alpha \beta} [\vec R_t] = \langle \alpha(t, t'') \vert \beta(t, t')
\rangle\;.
\label{eq:ampl}
\end{equation}
Next, because the Coulomb potential ${\cal U}_c(R_t)$ between A$_1$ and A$_2$
is a constant ($c$-number) in the quantum $\vec r$-space one can write down:
\begin{equation}
{\cal V}(R_t) = {\cal U}_c(R_t) + {\it W_{quant}}(R_t),
\label{eq:poten}
\end{equation}
where
\begin{equation}
{\it W_{quant}}(R_t) = {\it Re}\frac {\langle \alpha(t, t'')\vert
H(t)\vert \beta (t, t') \rangle}{T_{\alpha \beta}[\vec R_t]}.
\label{eq:poten2}
\end{equation}
The three-body hamiltonian is:
\begin{equation}
H(t) = -\frac{\hbar^2}{2m_n} \triangle_{\vec r} +
V_{13}(\vert \vec r - \vec R_1^t\vert) +
V_{23}(\vert \vec r - \vec R_2^t\vert)\;.
\end{equation}
Thus, the classical part of the three-body problem
(A$_1$, A$_2$, n) can be resolved in a self-consistent way. In practice, it can be realized,
for example, by few iterations: 1) for an arbitrary $R^{(0)}(t)$, e.g. straight-line trajectories,
we solve the quantum part of the problem. The
Eqs.(\ref{eq:maineqw}) - (\ref{eq:cond}) should be solved to get the
unknown amplitudes $C_n^1(t\rightarrow \infty)$ and
$C_n^2(t\rightarrow \infty)$ as time-dependent functions.
2) Now the effective potential ${\cal V}(\vec R(t))$
can be computed. To determine $R^{(1)}(t)$ one can
employ the expression \cite{newton65}
\begin{equation}
t = M \int_{r_m}^{R'} \frac{R dR}
{\sqrt {2MR^2 (E - {\cal V}^{(1)}(R)) - J^2}},
\end{equation}
where $J=\rho \sqrt {2ME}$; $\rho$ is an impact parameter; $E$
is a collision energy and $M$ is the reduced mass of the nuclear cores
$M = M_{A_1} M_{A_2} / (M_{A_1} + M_{A_2})$.
Within the next step one needs to compute
$R_j$ points as a function of time $t_j$. In order to obtain a smooth function
one can make a spline $R^{(1)}(t) = \sum_{\alpha} Z_{\alpha j} (t - t_j)^\alpha, t_j\leq t \leq t_{j+1}$ and
obtain a first approximation for $R^{(1)}(t)$, and so on (i=1, 2, 3,.., iterations).
The cross-section for the  reaction is \cite{pechukas69}, \cite{newton65}:
\begin{equation}
\left( \frac{d\sigma} {d\Omega} \right) =
\left( \frac{d\sigma} {d\Omega}\right)_{cl} \vert T_{\alpha \beta}[\vec R(t)]\vert^2,
\label{eq:cross}
\end{equation}
where
\begin{equation}
\left( \frac{d\sigma} {d\Omega} \right)_{cl} = \frac{\rho(\theta) \csc (\theta )}{\vert d\theta/d\rho \vert},
\end{equation}
and
\begin{equation}
\theta (\rho) = \pi - 2 \int_{r_m}^{\infty }\frac{dR}
{R^2 \sqrt{\rho^{-2}(1 - {\cal V}^{(I)}/E) - R^{-2}}}\;,
\label{eq:angle}
\end{equation}
here $r_m$ is a maximum of $R$ when the root is zero.
%
\section{Quotient analytical solution of the semiclassical Faddeev equations}
In this section we consider a special case of a neutron transfer reaction when the heavy nuclei
A$_1(Z_{A_1}, M_{A_1})$ and A$_2(Z_{A_2}, M_{A_2})$ are identical particles and then
we restrict ourselves to the two-level approximation in the
expansion (\ref{eq:expan}), i.e. $n=m=1$. To describe the matrix elements we delineate:
\begin{equation}
{\cal W}^{12}_{nm}(R(t),t) = {\cal W}^{21}_{mn}(R(t),t) = {\cal W}(R(t)),
\end{equation}
and the binding energies in the two channels are equal too: $E^{13}_{n=1}=E^{23}_{m=1}$.
So it turns out that the equations (\ref{eq:maineqw})-(\ref{eq:maineq}) can be solved in an explicit way:
\begin{equation}
C_1^{1}(\vec R(t), t) = C^{(A_1,n)}(\vec R(t), t) = \hbar \cos\left(\int_{-\infty}^t {\cal W}(R(t'))dt'\right ),
\label{eq:resolve1}
\end{equation}
\begin{equation}
C_1^{2}(\vec R(t), t) = C^{(A_2,n)}(\vec R(t), t) = {\rm i}\hbar \sin\left(\int_{-\infty}^t {\cal W}(R(t'))dt' \right).
\label{eq:resolve}
\end{equation}
Now, taking into account that:
\begin{equation}
dt = \frac{MRdR}{\sqrt{2MR^2(E-{\cal V}(R_t))-J^2}},
\end{equation}
where ${\cal V}(R_t) = {\cal U}_c(R_t) + {\it W_{quant}}(R_t)$,
${\cal U}_c(R_t) = (Z_{A_1}Z_{A_2}e^2)/R_t$,
$E$ is the c.m. collision energy, $M$ is the reduced mass, $e$ is the elementary charge.
Thus, the three-body transfer cross section can be written down as:
\begin{equation}
\left( \frac{d\sigma} {d\Omega} \right)_{tr} =
\left( \frac{d\sigma} {d\Omega} \right)_{cl}\hbar^2
\sin^2\left(2M\int_{r_m}^{+\infty}\frac{{\cal W}(R)RdR}
{\sqrt{2MR^2(E-{\cal V}(R_t))-2ME\rho^2(\theta)}}\right).
\label{eq:crosstr1}
\end{equation}
Here $\theta$ is the scattering angle of the classical particles \cite{newton65}:
\begin{equation}
\rho (\theta) = \frac{cot(\theta/2)}{2E},
\end{equation}
and also from \cite{newton65}:
\begin{equation}
\left( \frac{d\sigma} {d\Omega} \right)_{cl} = \left(
\frac{Z_{A_1}Z_{A_2}e^2}{4E\sin^2 (\theta /2)}\right)^2.
\end{equation}
Thereby our final result for the semiclassical three-body neutron transfer cross-section is:
\begin{eqnarray}
\sigma_{tr}(E)  =  2\pi \hbar^2 \int_{0}^{\pi}d\theta \sin\theta
\left(\frac{Z_{A_1}Z_{A_2}e^2}
{4E\sin^2 (\theta /2)}\right)^2\nonumber \\
\times \sin^2\left(2M\int_{r_m}^{+\infty}\frac{{\cal W}(R)
RdR}{\sqrt{2MR^2(E-{\cal V}(R_t))-2ME\rho^2(\theta)}}\right).
\label{eq:crosstr2}
\end{eqnarray}

Let us now proceed to a calculation of the effective
{"\sl quantum-classical"} potential ${\cal V}(R(t))$
between A$_1$ and A$_2$ in the transfer channel. To define amplitudes for the reaction
we have to adopt the limit $t\rightarrow \infty$
which is equivalent to $t=t^{\prime \prime}$ in Eq. (\ref{eq:ampl})
\begin{equation}
T_{\alpha \beta}[\vec R_t] = \langle \alpha \vert \beta (t^{\prime}, t^{\prime \prime})\rangle,
\end{equation}
here $\vert \alpha \rangle$ corresponds to the outgoing
(A$_2$,\ n)-bound state wave function
\begin{equation}
\vert \alpha \rangle = \vert \Phi_{\nu}^{(A_2, n)} (\vert \vec r -
\vec R_2(t)
\vert) \rangle \vert_{t\rightarrow \infty},
\end{equation}
where $\nu$ denotes a quantum state of the system (A$_2$,\ n), e.g. $\nu=1$. Next, 
$\vert \beta (t^{\prime}, t^{\prime \prime}) \rangle$ corresponds
to the total wave function of our three-particle system:
\begin{eqnarray}
\vert \beta(t^{\prime}, t^{\prime \prime}) \rangle = \vert \Psi (t)
\rangle \vert_{t\rightarrow \infty} \approx \Big [C^{(A_1,n)}(t)
\Big | \Phi_{\nu}^{(A_1,n)} (\vert \vec r - \vec R_1(t)\vert)\Bigr > \nonumber \\
+\ C^{(A_2,n)}(t)\Big | \Phi_{\nu}^{(A_2,n)} (\vert \vec r -
\vec R_2(t) \vert)\Bigr > \Big ]_{t\rightarrow \infty},
\end{eqnarray}
and for the nucleon transfer channel we have:
\begin{equation}
T_{\alpha \beta}[\vec R(t)] = C^{(A_2,n)}(\infty).
\end{equation}
Thus, the effective potential is:
\begin{eqnarray}
{\cal V}(R_t) &=& (Z_{A_1}Z_{A_2})e^2/R
+{\it Re}\Big (C^{(A_1,n)}(\infty)/C^{(A_2,n)}(\infty)
\{ \langle \Phi_{\nu}^{(A_2,n)}(\vec y)\vert p^2/2m\nonumber \\
& + & V_{13}(x)\vert \Phi_{\nu}^{(A_1,n)}(\vec x)\rangle
+ \langle \Phi_{\nu}^{(A_2,n)}(\vec y)\vert V_{23}(y)
\vert\Phi_{\nu}^{(A_1,n)}(\vec x) \rangle \}\nonumber \\
& + & \{ \langle \Phi_{\nu}^{(A_2,n)}(\vec y)\vert p^2/2m
+ V_{23}(y)\vert\Phi_{\nu}^{(A_2,n)}(\vec y)\rangle\nonumber \\
& + & \langle \Phi_{\nu}^{(A_2,n)}(\vec y)\vert V_{13}(x)
\vert\Phi_{\nu}^{(A_2,n)}(\vec y)\rangle \}\Big ),
\end{eqnarray}
where:
\begin{eqnarray}
\vec y = \vec r - \vec R_2(t),\ \ \vec x = \vec r - \vec R_1(t).
\end{eqnarray}

Finally:
\begin{eqnarray}
{\cal V}(R_t) &=& (Z_{A_1}Z_{A_2})e^2/R
+{\it Re}\Big [C^{(A_1,n)}(\infty)/
C^{(A_2,n)}(\infty)
\int d^3\vec r\left ( \Phi_{\nu}^{(A_2,n)}(\vec y)\right )^* V_{23}(y)\nonumber\\
&\times& \Phi_{\nu}^{(A_1,n)}(\vec x) + \int d^3\vec r\left (\Phi_{\nu}^{(A_2,n)}(\vec y) \right )^*
V_{13}(x)\Phi_{\nu}^{(A_2,n)}(\vec y)\Big ],
\end{eqnarray}
where amplitudes $C^{(A_1,n)}$ and $C^{(A_2,n)}$ are from Eqs. (\ref{eq:resolve1})-(\ref{eq:resolve}).

Now one can do the following: in the first step (or we could name it as a zero-th (0-th) approximation)
we only retain the Coulomb interaction in the effective potential
${\cal V}_{(0)}(R_t) = (Z_{A_1}Z_{A_2})e^2/R$,
and then calculate the 0-th approximation to the amplitudes
$C^{(A_1,n)}_{(0)}(\infty)$ and $C^{(A_2,n)}_{(0)}(\infty)$. In the
second step we can then compute the quantum corrections to the
effective potential ${\cal V}_{(1)}(R_t)$ then in the first approximation one can compute
the amplitudes $C^{(A_1,n)}_{(1)}(\infty)$ and $C^{(A_2,n)}_{(1)}(\infty)$
and of course one could continue the process if desired.

Let us write down the matrix element ${\cal W}(R(t))$ 
\begin{equation}
{\cal W}(R(t)) \enspace = \enspace \int d^3r (\varphi_{\nu}^{(A_j,n)}(\vec r - \vec R_j(t)))^*
V_{j3}(\vec r - \vec R_j(t)) \varphi_{\nu}^{(A_k,n)}(\vec r - \vec R_k(t)),
\label{eq:wmatel7}
\end{equation}
where $V_{j3}(x)$ is a local interaction,
e.g. a potential pit for the bound system $^{16}$O-n which gives
bound states and $\varphi_{\nu}^{(A_j,n)}(\vec r - \vec R_j(t))$ are
its wave functions. The results obtained can be used to describe
the one n-transfer reaction between two $^{16}$O nuclei.
This example has also been numerically calculated in
\cite{revai85} using single-term separable potentials and straight-line trajectories.
In the framework of the current formalism the simple expressions
(\ref{eq:crosstr1})-(\ref{eq:crosstr2}) and (\ref{eq:wmatel7}) have been derived by taking
into account the Coulomb potential between
nuclear cores and using the local nuclear potentials between A$_i$ and n $(i=1,2)$.
The elastic and transfer reaction cross-sections
are obtained using the self-consistent Pechukas method. In turn the Pechukas method
takes into account the interplay between the classical and
quantum degrees of freedom in the semiclassical
system and is consistent with the conservation laws of energy and angular
momentum \cite{pechukas69}.

It is essential to note here, that the same consideration as above
could be carried out for the three-body break-up channel. This is a very attractive and complicated
problem in the field of the heavy-ion collisions. Namely, a neutron emission reaction and/or
a charge particle, such as the $\alpha$-particle emission process.
In the case of such reactions, for instance
the $\alpha$-particle emission, in Eqs. (\ref{eq:break1})-(\ref{eq:emission}) we would need to apply
Coulomb asymptotic wave function in the three-body continuum. Obviously,
the effective potential between the heavy particles A$_1$ and A$_2$
will also be different in the three-body break-up channel.

\section{Conclusion}
We have formulated a semiclassical approach for a model
three-body system with two heavy nuclear cores
A$_1$ and A$_2$ moving along classical trajectories
and a lighter particle n, i.e. neutron. The three-body system is shown in Fig. 1.
The quantum dynamics of n is described based on the
few-body quantum-mechanical Eqs. (\ref{eq:fadd11})-(\ref{eq:fadd})
with realistic (local) nuclear-nuclear potentials $V_{13}(\vec x)$ and $V_{23}(\vec y)$.
The classical dynamics of A$_1$ and A$_2$ are described based on
Newtonian (non-relativistic) mechanics Eq. (\ref{eq:class}).
However, this becomes important, with the use of the Pechukas self-consistent method we could
take into account the interplay between classical
and quantum degrees of freedom in the system and thereby obtain even more realistic trajectories
for the classical particles. Therefore, the proposed method
is divided into two parts: the 1st part is the quantum-mechanical problem for a lighter particle "n"
dipped into the nuclear potential pits of the heavy particles A$_1$ and A$_2$, the second
part is the classical problem for two heavy nuclear cores interacting by the quantum and classical (Coulomb)
self-consistent potential ${\cal V}(\vec R(t))$, i.e. Eqs. (\ref{eq:poten})-(\ref{eq:poten2}).
Also, it would be appropriate to make few comments about the semiclassical
Faddeev-type equations, i.e. Eqs. (\ref{eq:fadd11})-(\ref{eq:fadd}).
First of all, the constructed coupled equations satisfy the Schr\" odinger equation exactly. Secondly,
the Faddeev decomposition avoids the over-completeness problems. Therefore,
two-body subsystems are treated in an equivalent way and the correct asymptotic is guaranteed
\cite{fb_my}. The current method
simplifies the solution procedure and provides the correct asymptotic behavior of the 
solution. 
Finally, the Faddeev-type equations have the same advantages as the original Faddeev 
equations, because they are formulated for the three-body wave function components	$\Psi_1(\vec r, \vec R(t), t)$
and $\Psi_2(\vec r, \vec R(t), t)$ with correct physical asymptotes.

In the solution of the time-dependent Eqs.
(\ref{eq:fadd11})-(\ref{eq:fadd}) one needs to consider the number of
channels $n$ which are needed to be included
in the close-coupling expansion (\ref{eq:expan}). This is an important issue, because $n$
controls the number of coupled differential equations to be numerically solved, i.e. Eqs. (\ref{eq:maineqw})-(\ref{eq:maineq}).
However, in the actual numerical computation
one could only retain a few states in Eq. (\ref{eq:expan}). For example, it is quite reasonable to expect that for
closed shell nuclei, e.g. A$_i \equiv ^4$He,$^{12}$C or $^{16}$O,
just one or two states should be predominate during low energy collisions.
Next, the expression (\ref{eq:trans}) is true for an inertial coordinate system,
i.e. when $v=\dot {\vec R}(t)$=const.
In the case of the realistic trajectories $\dot {\vec R}_i(t) \neq$const and
one needs to make considerable alterations in the expression and in the theory.
However, at low energies when $v\approx 0$   
the exponent multiplier is approximately equal to one, i.e.
$
e^{im_n \vec v \vec r-i\frac{m_n}{2}v^2t} \approx 1.
$

In conclusion, as mentioned in the introduction, few-body semiclassical models in
nuclear physics can help us gain deeper insight into complex nuclear processes. Specifically,
in the case of identical heavy nuclei and a two-level approximation in the expansion (\ref{eq:expan})
the resulting set of coupled differential Eqs. (\ref{eq:maineqw})-(\ref{eq:maineq})
can be resolved analytically, i.e. expressions (\ref{eq:resolve1})-(\ref{eq:resolve}). This analytical solution
might be useful, for example, in the investigation of the nucleus
$^{13}$C, e.g. in the collision $^{13}$C + $^{12}$C $\rightarrow$ $^{12}$C + $^{13}$C; $^{12}$C + $^{12}$C + n.
The structure of $^{13}$C=$(^{12}\mbox{C},n)$ can be
important for low energy reactions in the s-process
neutron source in stars $^{13}\mbox{C}(\alpha,n)^{16}\mbox{O}$, see for example
\cite{hollo89,brune93,lacognata2012}. Also, we would like to note, that a possible relativistic expansion
of the semiclassical theory presented above in this paper would be a very useful future work.

\acknowledgments{
This work was partially supported by the Office of Sponsored Programs and
Integrated Science and Engineering Laboratory Facility {\it(ISELF)} of
St. Cloud State University, St. Cloud, Minnesota 56301-4498, USA.}


\clearpage
\begin{figure}
\begin{center}
{
\unitlength 1.00mm
\begin{picture}(110,190) (10,-20)
\thicklines
\put(25,115){\circle{14}}      
\put(25,125){\makebox(0,0)[cc]{A$_1$}}
\put(25,115){\makebox(0,0)[cc]{$M_1, Z_1^+$}}
\put(102,55){\circle{14}}      
\put(102,55){\makebox(0,0)[cc]{$M_2, Z_2^+$}}
\put(104,44){\makebox(0,0)[cc]{A$_2$}}

\linethickness{0.001mm}
\put(7,115){\line(1,0){11}} 
\put(32,115){\line(1,0){87}}

\put(109,55){\line(1,0){9}} 
\put(7,55){\line(1,0){88}} 
\linethickness{0.1mm}

\put(32,115){\vector(1,0){11}} 
\put(95,55){\vector(-1,0){17}} 

\put(12,55){\vector(0,1){60}} 
\put(15,100){\makebox(0,0)[cc]{\LARGE $\vec \rho$}}

\put(97,60){\vector(-4,3){66.7}} 

\put(70,80.3){\circle*{1}}      
\put(73,83){\makebox(0,0)[cc]{\LARGE $O$}}
\put(70,80.3){\vector(-1,2){27.2}} 

\put(42,136){\circle*{3}}      
\put(49,130){\makebox(0,0)[cc]{\LARGE $\vec r$}}  
\put(45,87){\makebox(0,0)[cc]{\LARGE $\vec R_{12}(t)$}}

\put(40,120){\makebox(0,0)[cc]{\LARGE $\vec v_1$}}  
\put(110,128){\makebox(0,0)[cc]{\LARGE $\vec v^{\ \prime}_1$}}  

\put(82,51){\makebox(0,0)[cc]{\LARGE $\vec v_2$}}  
\put(20,45){\makebox(0,0)[cc]{\LARGE $\vec v^{\ \prime}_2$}}  

\put(40,140){\makebox(0,0)[cc]{\LARGE n}}  

\linethickness{0.7mm}
\bezier{33}(55,115)(95,117)(118,125)    
\put(117.5,124.5){\vector(3,2){2}} 
\bezier{33}(75,55)(20,53)(7,45)    
\put(7,45){\vector(-3,-2){2}} 


\end{picture}
\vspace{-5cm}
\caption{Neutron (n) few-body
quantum dynamics between two classically moving nuclear cores: A$_1$=($M_1, Z^+_1$)
and A$_1$=($M_2, Z^+_2$). Here, $M_i, Z^+_i$ are masses and Coulomb charges respectively
of A$_i$ ($i=1,2$), $O$ is the center-of-mass of the three body system,
$\vec r$ is the coordinate radius-vector of n, $\vec R_{12}(t)=\vec R_1(t)-\vec R_2(t)$
is the separation vector between ($M_1, Z^+_1$) and ($M_2, Z^+_2$),
$\vec R_1(t)$ and $\vec R_2(t)$ are the radius-vectors of A$_1$ and A$_2$
($\vec R_{1(2)}(t)$ are not shown in this figure), $t$ is the time in the system,
$\vec v_1$  and $\vec v_2$ are initial at $t\rightarrow -\infty$ velocities of
($M_1, Z^+_1$) and ($M_2, Z^+_2$) respectively, $\vec v^{\ \prime}_1$
and $\vec v^{\ \prime}_2$ are final at $t\rightarrow +\infty$
velocities of ($M_1, Z^+_1$) and ($M_2, Z^+_2$) respectively,
$\rho$ is the impact parameter of the three-body collision:
$[(M_1, Z^+_1), \mbox{n}] + (M_2, Z^+_2) \rightarrow [(M_2, Z^+_2), \mbox{n}] + (M_1, Z^+_1);$
$(M_2, Z^+_2) + (M_1, Z^+_1) + \mbox{n}$, where the nucleon transfer from ($M_1, Z^+_1$) to ($M_2, Z^+_2$)
and the three-body break-up channels are presented here. The nuclear interaction between n and A$_i\ (i=1,2)$
depends on the distances $|\vec x|$ and $|\vec y|$ between n and A$_1$ and between n and A$_2$ 
respectively ($\vec x$ and $\vec y$ are not shown in this figure).}
\label{fig1}
}\end{center}
\end{figure}
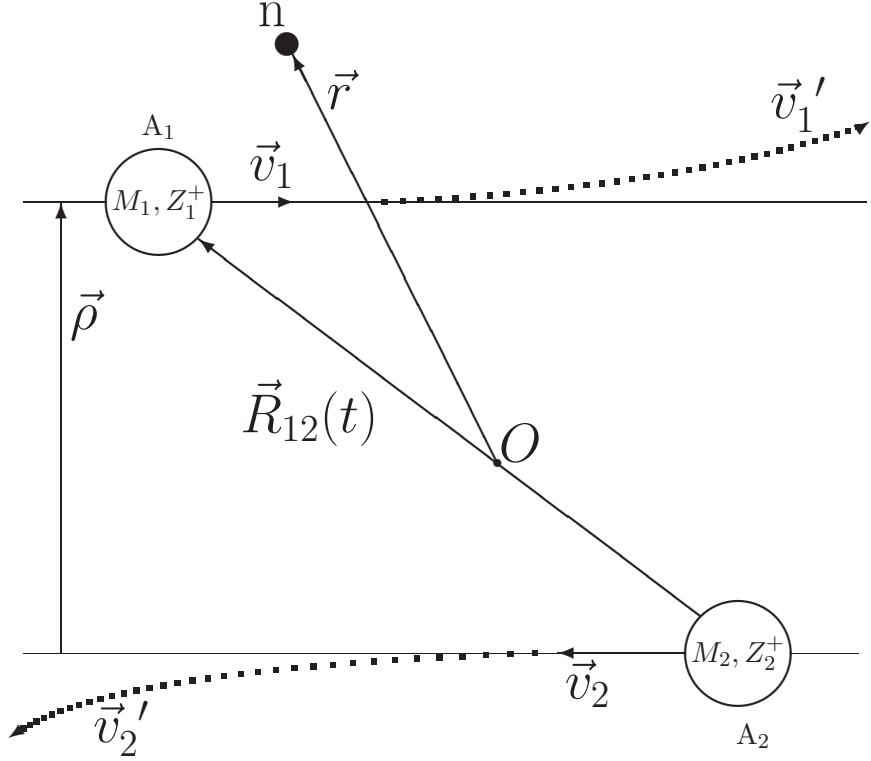
\end{document}